\documentclass[conference]{IEEEtran}
\IEEEoverridecommandlockouts
\usepackage{cite}
\usepackage{amsmath,amssymb,amsfonts}
\usepackage{algorithmic}
\usepackage{soul}
\usepackage{graphicx}
\usepackage{textcomp}
\usepackage{physics}
\usepackage{xcolor}
\usepackage[margin=0.724in]{geometry}
    
\def\BibTeX{{\rm B\kern-.05em{\sc i\kern-.025em b}\kern-.08em
    T\kern-.1667em\lower.7ex\hbox{E}\kern-.125emX}}

\usepackage{subfig}
\usepackage{hyperref}

\def\BibTeX{{\rm B\kern-.05em{\sc i\kern-.025em b}\kern-.08em
    T\kern-.1667em\lower.7ex\hbox{E}\kern-.125emX}}
    
\DeclareCaptionLabelFormat{andtable}{#1~#2  \&  \tablename~\thetable}
\IEEEoverridecommandlockouts

\title{Resiliency Analysis and Improvement of Variational Quantum Factoring in Superconducting Qubit}

\author{\IEEEauthorblockN{  Ling Qiu, Mahabubul Alam, Abdullah Ash-Saki, Swaroop Ghosh}
\IEEEauthorblockA{Department of Electrical Engineering\\
Pennsylvania State University, University Park, PA-16802 \\
lingq@psu.edu, mxa890@psu.edu, axs1251@psu.edu, szg212@psu.edu}

}

\begin{document}

\maketitle

\begin{abstract}
Variational algorithm using Quantum Approximate Optimization Algorithm (QAOA) can solve the prime factorization problem in near-term noisy quantum computers. Conventional Variational Quantum Factoring (VQF) requires a large number of 2-qubit gates (especially for factoring a large number) resulting in deep circuits. The output quality of the deep quantum circuit is degraded due to errors limiting the computational power of quantum computing. In this paper, we explore various transformations to optimize the QAOA circuit for integer factorization. We propose two criteria to select the optimal quantum circuit that can improve the noise resiliency of VQF.  

\end{abstract}

\section{{Introduction}}
Quantum computing is prophesized to solve integer factorization which is the basis of the Rivest–Shamir–Adleman (RSA) based cryptosystem. 
Quantum factoring using Shor's algorithm has demonstrated the potential to factor large integers in polynomial time compared to classical computers which take exponential time. However, it requires an excessive number of qubits to factor even trivial numbers e.g. 21 precluding its application in today's Noisy-Intermediate-Scale-Quantum (NISQ) computers that possess a limited number of qubits \cite{vandersypen2001experimental}. To make the best use of the limited quantum resources, an alternative approach is to transform the factoring problem into a combinatorial optimization problem which is then solved using a hybrid quantum-classical solver known as Variational Quantum Factoring (VQF) \cite{anschuetz2019variational}. 

\begin{figure*} [t]
\vspace{-1em}
 \begin{center}
    \includegraphics[width=0.95\textwidth]{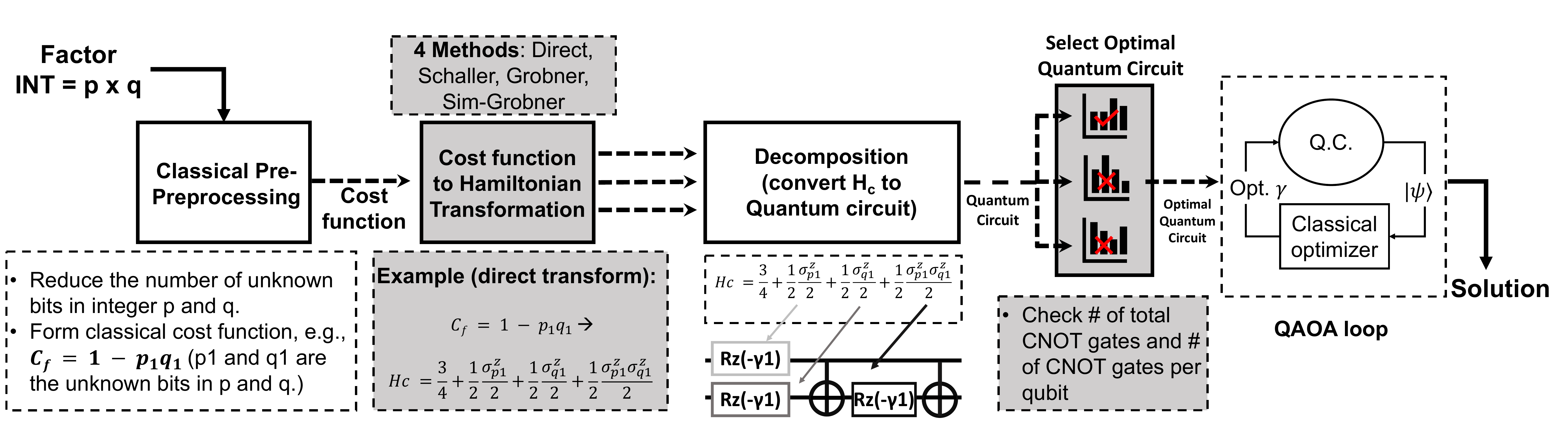}
 \end{center}
 \vspace{-1em}
 \caption{The proposed flow of VQF (the grey boxes are the added stages to the conventional VQF.) } \label{fig:flow}
\end{figure*}

 The abstract flow of VQF is shown in Fig. \ref{fig:flow}. In the first stage, we formulate the factoring problem as a cost function with binary variables. With the Boolean variable properties, we perform classical pre-processing to reduce the number of variables in the cost function which is then encoded into a cost Hamiltonian such that its ground state encodes the optimal solution to the optimization problem. In the next stage, the cost Hamiltonian is decomposed into a Parameterized Quantum Circuit (PQC) (i.e. a quantum circuit consisting of parameterized gates). In the final stage, the quantum circuit is passed to a hybrid quantum-classical solver to optimize its parameters iteratively to minimize the cost function. The hybrid optimization is terminated when a pre-defined optimization goal is satisfied. In this work, we use the Quantum Approximate Optimization Algorithm (QAOA), one of the promising hybrid quantum-classical solver, to solve the optimization problem \cite{farhi2014quantum}.   
 
In the NISQ era quantum computers, the performance of VQF can be affected by the quantum noises \cite{alam2019analysis, zhou2018quantum} e.g., gate error and decoherence. Gate error is the imprecision of applying a quantum gate whereas decoherence noise is rooted in the qubits' loss of information due to the interaction with the environment. Quantum noises can deviate the modulation of a quantum state from its original planned path thus affecting the VQF performance. The impact of decoherence error on similar quantum optimization problems have already been well-studied and known to be solely dependent on the circuit depth \cite{alam2019analysis}. Therefore, we focus on gate error on VQF.

Factoring a large number may require high-depth QAOA (i.e., higher $p$) to improve the VQF performance. However, we show that the noise resiliency of VQF drops at higher $p$-levels due to noise. Therefore, it is important to study the behavior of VQF under noise and develop techniques to improve resiliency. We have noted that a cost Hamiltonian can be mathematically transformed into different forms, which are then decomposed to different quantum circuits. The transformed quantum circuit flavors can offer a varied degree of noise resilience. 

\emph{To our best knowledge, this is the first work to quantify the impact of noise on VQF and improve its performance using various transformations}. 
We make the following contributions:

\noindent \textbf{(a) Study the impact of gate error on VQF:} We show that gate error significantly impacts the VQF performance.  

\noindent \textbf{(b) Analyze the impact of transformations on VQF:} We study the noise resilience of VQF with the resulting quantum circuits derived from $4$ transformations. 

\noindent \textbf{(c) Evaluation of integers:} We study 2 realistic integers with prime factors and 2 synthetic integers with non-prime factors. 

\noindent \textbf{(d) Propose a novel VQF implementation flow:} We integrate our criterion of selecting the quantum circuit into the workflow of VQF to improve its performance.

\textbf{Paper organization:} Section II introduces background on QAOA  and conversion of factoring problem into optimization problem. 
Section III explains the process and performance analysis of various transformations. Section VI presents future work and limitations. Conclusions are drawn in Section V.

\section{Preliminaries}

\subsection{Quantum Approximate Optimization Algorithm}
The optimal solution of a combinatorial optimization problem is obtained from a finite set of solutions. The unknowns are on an N-bit binary strings: $Z = \{ z_1, z_2, ... z_n$ \}, where $z_i \in \{0,1\}$. The goal of this optimization problem is to find a string to either maximize or minimize (depending on the problem) the cost function $C(Z)$. In VQF, we seek to minimize $C(Z)$ which consists of $m$ clauses, each describing a constraint to the optimization problem. It is denoted as: $C(Z) = \sum\limits_{i=1}^m C_m(Z)$.

\begin{figure} [t] 
\vspace{-1em}
 \begin{center}
    \includegraphics[width=0.495\textwidth]{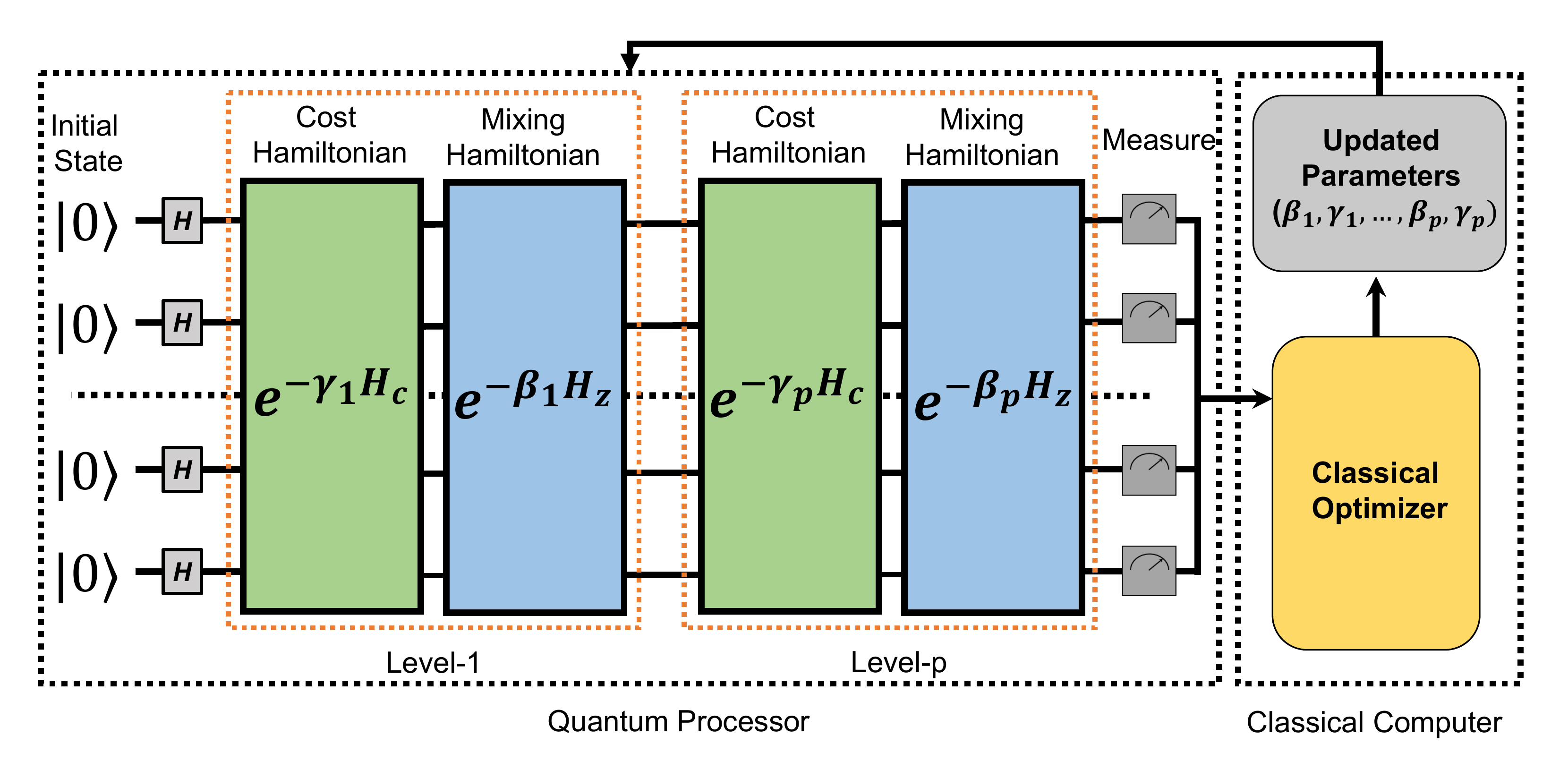}
 \end{center}
 \vspace{-1em}
 \caption{Schematic diagram of $p$-level QAOA. } \label{fig:qaoa}
 \vspace{-3mm}
\end{figure}

QAOA is a promising algorithm to tackle the combinatorial optimization problem in the NISQ era due to its inherent error resiliency. An overview of a p-level QAOA is shown in Fig. \ref{fig:qaoa}. The initial state is set by applying a Hadamard gate to each of the qubits, s.t. $\psi_I = \ket{+}^{\otimes n}$, where $n$ is the number of qubits, $\ket{+} = \frac{1}{\sqrt{2}}(\ket{0}+\ket{1})$ and ${\otimes}$ stands for the tensor product. Each unknown in the cost function is mapped to a qubit. For gate-based quantum computers, the cost Hamiltonian and mixing Hamiltonian are decomposed into the quantum circuits with the native gates of the target hardware, which are applied repeatedly for $p$ times to generate the output state, $\psi_F$. We explain the decomposition technique in the next section. For VQF, the rotational parameters are optimized iteratively by minimizing the expectation value of $\psi_F$ via classical optimization. A lower expectation value indicates that $\psi_F$ is closer to the objective state, and the solution binary string can be retrieved with a higher probability by repetitive measurement. It is well-known that the QAOA performance improves with higher $p$ in the noiseless environment. However, we show that in reality, the performance may degrade with an increasing $p$ due to quantum noises.

\subsection{Factoring as Binary Optimization}
\begin{figure} [hb] 
\vspace{-1em}
 \begin{center}
    \includegraphics[width=0.45\textwidth]{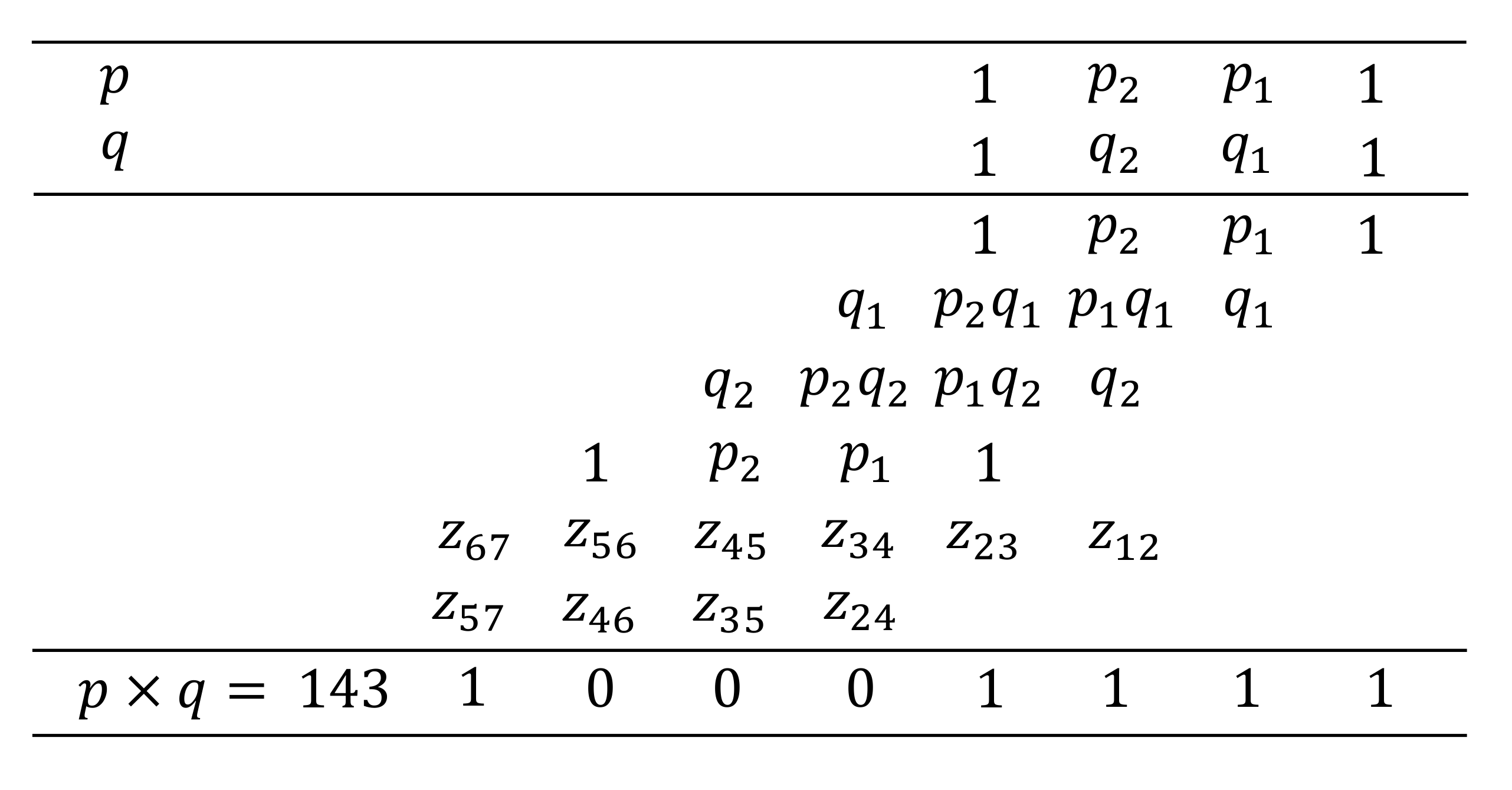}
 \end{center}
 \vspace{-1em}
 \caption{Multiplication table of 143 in binary representation. } \label{fig:mult-table}
 \vspace{-3mm}
\end{figure}

It has been proved that a factorization problem can be mapped into an optimization problem \cite{burges2002factoring}. For demonstration purposes, we adopt the idea from \cite{xu2012quantum} that focuses on a specific factorization, where both of the multiplier's bit length are the same e.g., factoring of $143 = 11*13$ to describe the VQF procedure. Since $143$ is an odd number, the least significant bit (LSB) for both multipliers are $1$. The multiplication table is constructed as shown in Fig. \ref{fig:mult-table}. Here $p$ and $q$ are the multipliers represented in binary; $p_i$ and $z_{ij}$ stands for the $i$-th bit of $p$ and the carry from the $j$-th bit to the $i$-th bit, respectively; The last row is $143$ in binary. Given the previous assumptions of this factorization, the unknown variables in this multiplication tables are: $p_2$, $p_1$, $q_2$, $q_1$ and the carry bits. 

The equations that describe the multiplication relations column-wise are written as:
\begin{gather}
p_1 + q_1 = 1+2z_{12} \\
p_2 + p_1q_1 + q_2 + z_{12} = 1+2z_{23}+4z_{24} \\
... \\
1 + z_{56} + z_{46} = 0+2z_{67} \\
z_{67}+z_{57} = 1
\end{gather}
\label{eq: 6}



we simplify the above equations based on the Boolean properties to reduce the number of qubits. This process is denoted as the classical pre-processing. For example, in equation (1), we note that $z_{12}$ has to be $0$ given that the maximum result on the left-hand side, $p_1 + q_1$, is $2$. Similarly, we can simplify the rest of the equations. The equations after the classical pre-processing are shown below:
\begin{gather}
p_1 + q_1 - 1 = 0 \\
p_{2}+q_{2} - 1 = 0 \\
p_{2}q_{1} + p_{1}q_{2} - 1 = 0
\end{gather}
The corresponding cost function, denoted as $C_f$, is constructed by the cumulative square sum of the above equations. i.e.:
\begin{equation}
\begin{aligned}
C_f = &(p_1 + q_1 - 1)^2 + (p_{2}+q_{2} - 1)^2 + \\
&(p_{2}q_{1} + p_{1}q_{2} - 1)^2 
\end{aligned}
\label{eq: 6}
\end{equation}

\subsection{Mapping Cost Function into Cost Hamiltonian}
To construct the Hamiltonian model, each of the variables ($z_i$) in the binary string is mapped into a quantum spin, $\sigma_{i}^{z}$, where $\sigma_{i}^{z} \in \{-1,+1\}$, such that: $\hat{z_i} = \frac{1-\sigma_{i}^{z}}{2}$.

Thus, the cost Hamiltonian for factoring $143$ shown in equation (\ref{eq: 6}) is constructed as:
\begin{equation}
\begin{aligned}
H_c = &(\hat{p_1} + \hat{q_1} - 1)^2 + (\hat{p_{2}}+\hat{q_{2}} - 1)^2 + \\
&(\hat{p_{2}}\hat{q_{1}} + \hat{p_{1}}\hat{q_{2}} - 1)^2 
\end{aligned}
\label{eq: 7}
\end{equation}

\section{Transformation Schemes and their Performance Analysis}

\subsection{Transformation Schemes}
There are multiple schemes to mathematically transform the cost Hamiltonian in equation (10) into other forms to optimize the resulting quantum circuit. 
In the following subsections, we introduce the 4 transformations namely, DIRECT \cite{anschuetz2019variational}, SCHALLER \cite{xu2012quantum}, GROBNER \cite{dridi2017prime} and SIM-GROBNER (proposed in this paper). 

\subsubsection{DIRECT}
\begin{figure} [hb] 
\vspace{-1em}
 \begin{center}
    \includegraphics[width=0.48\textwidth]{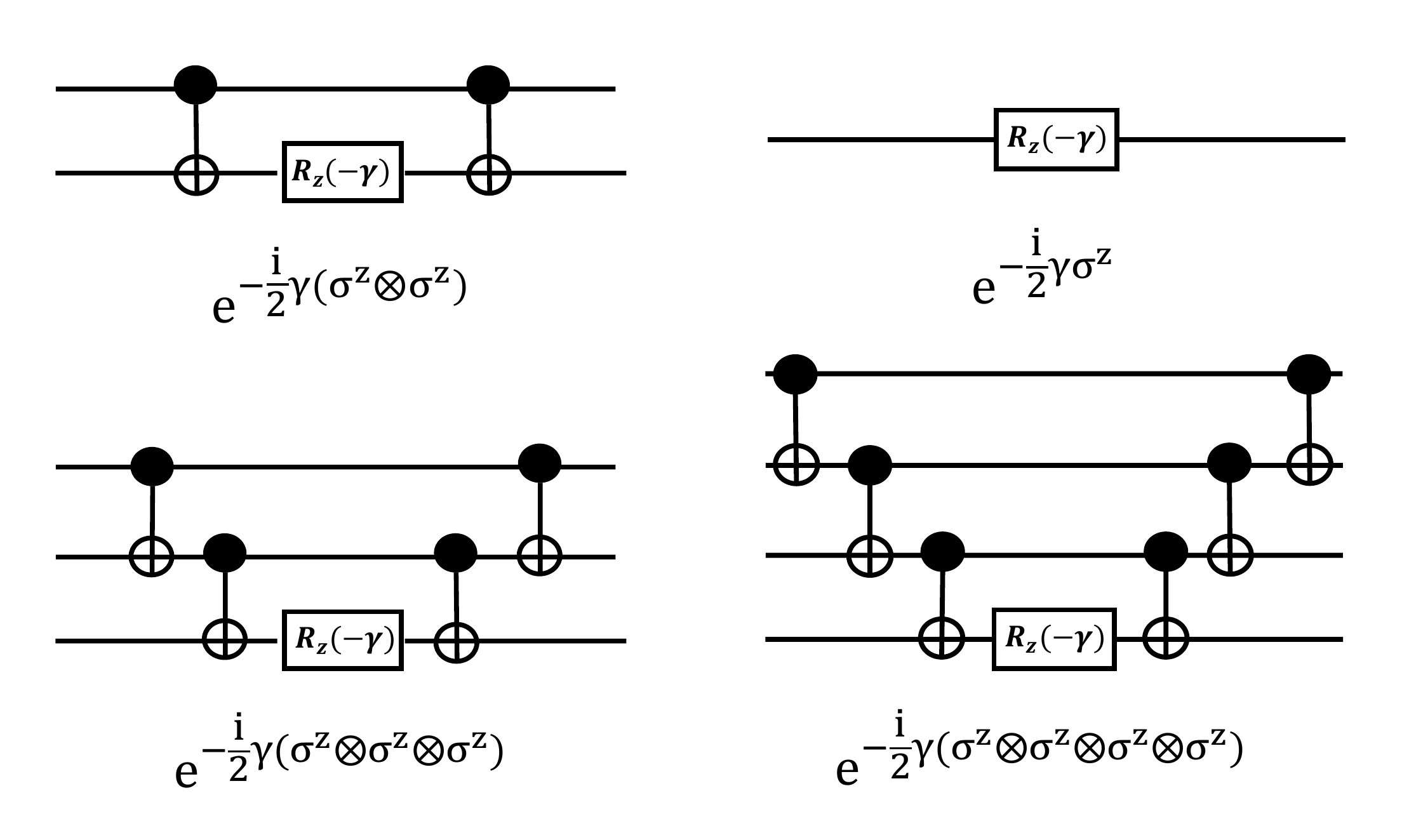}
 \end{center}
 \vspace{-1em}
 \caption{The circuit decomposition mapping. } \label{fig:structure}
 \vspace{-3mm}
\end{figure}

This transformation is carried out by directly expanding the original cost Hamiltonian (equation (\ref{eq: 7})) without any mathematical transformation. After the expansion, the cost Hamiltonian (for factoring 143) is shown in equation (\ref{eq: 88}): 

\begin{equation}
\begin{aligned}
H_c = &3 - \hat{p_1} - \hat{p_2} - \hat{q_1} - \hat{q_2} + \hat{2p_1}\hat{q_1} - \hat{p_2}\hat{q_1} + \\
& \hat{2p_2}\hat{q_2} + 2\hat{p_1}\hat{p_2}\hat{q_1}\hat{q_2} 
\end{aligned}
\label{eq: 88}
\end{equation}
The qubit interaction terms such as, $\sigma^{z}_{1}\sigma^{z}_{2}$ denotes $\sigma^{z}_{1}\otimes \sigma^{z}_{2}$ and so on. The circuit components are shown in Fig. \ref{fig:structure}. After decomposing the cost Hamiltonian into a quantum circuit with the basis gates e.g., CNOT, RX, RZ, and RY (native to IBM quantum computers), it can be noted that the maximum number of qubit interaction for DIRECT is $4$, which requires $6$ CNOT gates for construction.  

\subsubsection{SCHALLER}

The maximum number of qubit interaction can be reduced to $3$ using Schaller and Sch\"{u}tzhold transformation that constructs equation like $AB+S=0$ into a cost Hamiltonian of  $2[\frac{1}{2}(A+B-\frac{1}{2})+S]^2- \frac{1}{8}$. For factoring $143$, $(\hat{p_{2}}\hat{q_{1}} + \hat{p_{1}}\hat{p_{2}} -1)^2$ can be transformed into $2[\frac{1}{2}(\hat{p_1}+\hat{q_2}-\frac{1}{2})+\hat{p_2}\hat{q_1}-1]^2- \frac{1}{8}$. It can be noted in equation (\ref{eq: schaller}) that the maximum number of qubit interaction is reduced to $3$:
\begin{equation}
\begin{aligned}
H_c = &5 - 3\hat{p_1}-\hat{p_2}-\hat{q_1}+2\hat{p_1}\hat{q_1}-3\hat{p_2}\hat{q_1}+  \\ 
&2\hat{p_1}\hat{p_2}\hat{q_1}-3\hat{q_2}+\hat{p_1}\hat{q_2}+2\hat{p_2}\hat{q_2}+2\hat{p_2}\hat{q_1}\hat{p_2}
\end{aligned}
\label{eq: schaller}
\end{equation}


\subsubsection{GROBNER}

This transformation reduces the maximum level of qubit interaction from $4$ to $2$ by replacing a multiplication term of two variables with a new variable and adding a corresponding penalty term, $(p_iq_j - w_{ij})^{+}$ to $H_c$. Hence, this transformation decreases the level of qubit interaction at the cost of increased number of qubits (i.e. by adding extra unknown variables to the cost Hamiltonian). The penalty term, $(p_iq_j - w_{ij})^{+}$, can be obtained via Gr\"{o}bner bases computation:
\begin{equation}
\begin{aligned}
(p_iq_j - w_{ij})^{+} =  &a(p_iw_{ij}-w_{ij}) + b(q_iw_{ij}-w_{ij}) +  \\
& c(p_iq_j-w_{ij})
\end{aligned}
\label{eq: grobner}
\end{equation}
 where $a,b,c \in \mathbb{R}$ such that $-a-b-c > 0$, $-b-c > 0$, $-a-c>0$ and $c>0$. In this case, we take $a$, $b$ and $c$ to be $-2$, $-2$ and $1$, respectively. 
 

\subsubsection{SIM-GROBNER} We propose SIM-GROBNER transformation which also uses a replacement strategy but with a simplified penalty compared to GROBNER i.e.:
\begin{equation}
\begin{aligned}
(p_iq_j - w_{ij})^{+} &=  (p_iq_j - w_{ij})^2
\end{aligned}
\label{eq: logical 9}
\end{equation}

This scheme can reduce the maximum number of qubit interaction from $4$ to $3$ but will also increase the number of qubits.

\subsection{Experimental Setup}
\subsubsection{Studied Cases} The analysis is performed for factoring $143$ and $291311$ together with two other random cost Hamiltonians. Ideally, we should analyze the trends for a larger set of realistic numbers that are used in cryptography. However, such numbers may require a large number of qubits that cannot be supported by the present NISQ computers. Since our objective is to enhance the resilience of VQF to solve the factoring problem, we note that other random numbers (with non-prime factors) can also be used for the study.
Since there are only limited mathematical forms for the cost function of VQF, the selected $4$ studied cases are able to cover most of their characteristics. Therefore, the findings from these studied cases can be applied to factor realistic and large numbers when NISQ computers with a higher number of qubits are available. The random cost Hamiltonians are chosen such that their minimum values are equal to $0$ (similar to the cost Hamiltonians for actual prime factorization problems).

\subsubsection{Modeled Quantum Computer} To focus solely on the impact of various transformation techniques, we ignore the coupling constraints that are hardware-specific and prohibit some qubit to interact directly (and may require SWAP gates). We consider the implementation on a 16-qubit fully connected quantum computer such that two-qubit interactions (CNOT) is allowed between any two qubits. This quantum computer is modeled using IBM Qiskit \cite{Qiskit}. Without the loss of generality, we assume that each qubit has identical quality metrics. This further reduces the influence of the different configurations of quantum computer hardware. The values of gate errors are obtained by averaging the real gate error metrics across all qubits that are reported in the IBM 16-qubit Melbourne quantum computer on a randomly chosen day. Lastly, we scale the noise levels (from 0-100\%) obtained from the real quantum hardware to evaluate the sensitivity of VQF to noise. The implication of above assumptions are discussed in Section IV.

\subsubsection{Choice of Classical Optimizer} A variety of classical optimizers can be used with QAOA. The performance of local optimizers may greatly depend on the distribution of initial points. In addition, they may get trapped in a local optimum easily. To reduce the influence of the above factors, we use a global optimizer namely, differential evolution to train the QAOA circuit implemented from the Scipy-optimize library\cite{2019arXiv190710121V}. We constrain the search space to $\gamma_i \in [0,2\pi]$ and $\beta_i \in [0,2\pi]$ to boost the optimization speed.

\subsection{Obtaining Expectation Value}
The expectation value can be obtained by: $\langle E \rangle = \bra{\psi_F} H_c \ket{\psi_F}$. However, a $n$-qubit QAOA corresponds to an $n\times n$ matrix for $H_c$. Therefore, the size of $H_c$ increases exponentially as the number of qubit increases requiring an enormous amount of computing resources. 
We adopted the idea from \cite{zhou2018quantum} that uses Monte-Carlo simulations to obtain an approximated expectation value classically by averaging $M$ measurements:
\begin{equation}
\begin{aligned}
\Tilde{E} = \frac{1}{M}\sum\limits_{j=1}^MC(Z_j)
\end{aligned}
\label{eq: expectation}
\end{equation}
where $Z_j$ is the binary strings of the $j$-th measurement of the $\psi_F$ in the computational basis. Ideally, the approximation gets better with larger $M$ value. Empirically, we have discovered that a relatively large number of measurements is sufficient for a decent approximation but it is also related to the number of qubit.

\subsection{Circuit-level Comparison}

\begin{figure} [] 
\vspace{-1em}
 \begin{center}
    \includegraphics[width=0.5\textwidth]{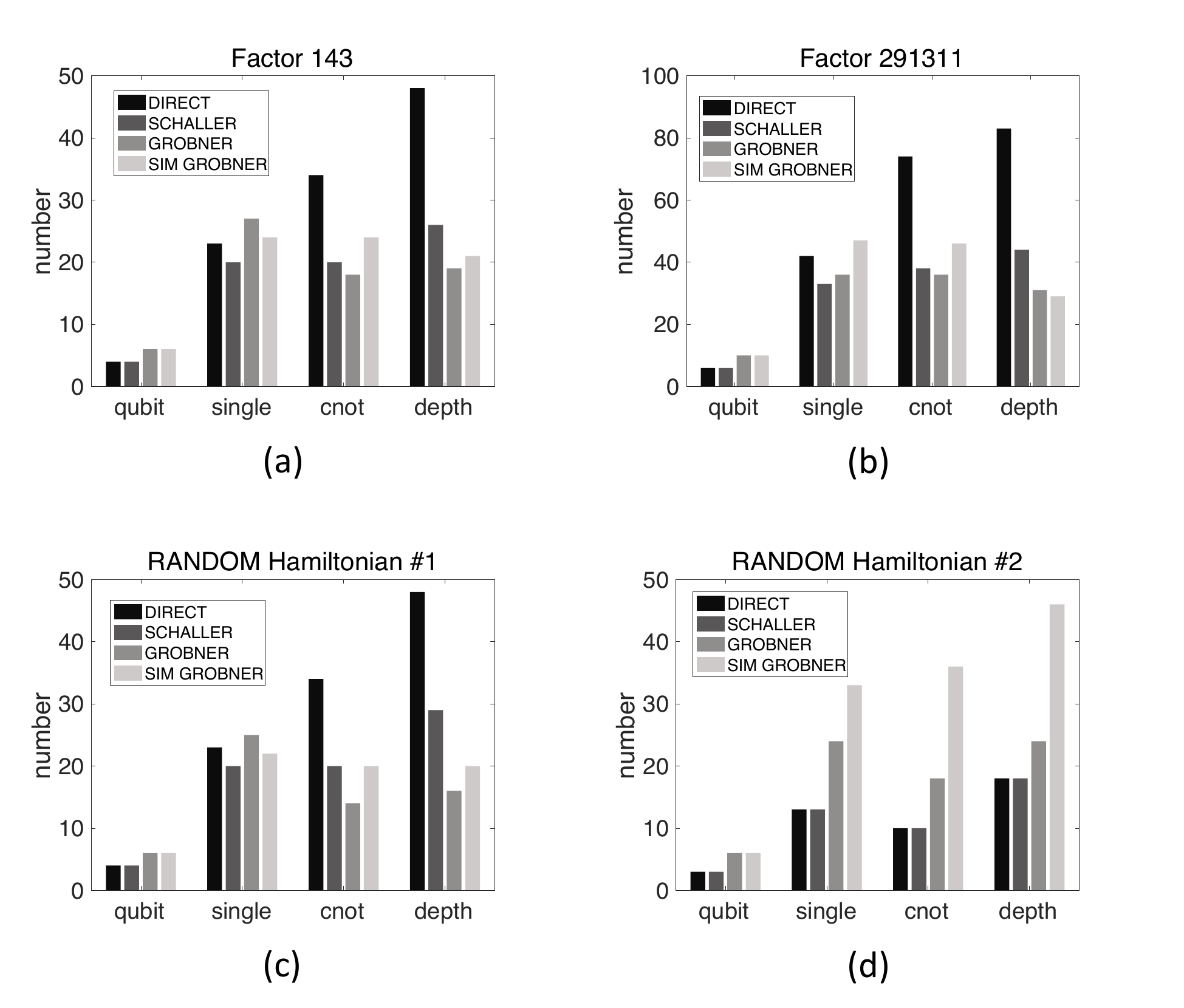}
 \end{center}
 \vspace{-1em}

 \caption{Comparison of the number of qubits, single gates, CNOT gates and circuit depth for VQF at $p$=$1$ for factorization of, (a) 143; (b) 291311; (c) random Hamiltonian \#1; (d) random Hamiltonian \#2.} \label{fig:arch}
 \vspace{-4mm}
\end{figure}

It is known that the resiliency of a quantum circuit depends on the number of qubits, noisy gate operations and circuit depth. Therefore, it is important to compare these key characteristics for various transformations as shown in Fig.\ref{fig:arch}.

For each number, GROBNER and SIM-GROBNER use the same (and higher) number of qubits than DIRECT and SCHALLER. This is due to the introduction of new variables into circuits to reduce the order of qubit interaction. In terms of the number of single-qubit gates, we did not note a consistent trend among the transformations. For the first three numbers (i.e. $143$, $291311$ and random Hamiltonian \#1), DIRECT has a significantly higher number of CNOT operations. For example, it has $60\%$ more CNOT gates than GROBNER for the factorization of $291311$. The reason is $4$-qubit interactions for these three cases each of which costs as many as $6$ CNOT gates (shown in Fig.\ref{fig:structure}). For the random Hamiltonian \#2, DIRECT's highest qubit interaction is $3$. Therefore, applying SCHALLER transformation results in the same circuit architecture and applying SIM-GROBNER becomes unnecessary since it increases both the number of qubits, CNOT gates and circuit depth. This study indicates two things: first, the transformations offer a trade-off space among the number of qubits, circuit depth and CNOT gates, and, second, the choice of appropriate transformation is dependent on the target integer for factorization (i.e., cost Hamiltonian).

\subsection{Evaluation Metric} 

To quantify the performance gain of VQF under noise across various factoring problems, we define the normalized residual performance gain as $G_{i,p}$ (denoted as NRPG in figures), where $i$ and $p$ is the noise level and QAOA level, respectively. It is calculated as:
\begin{equation}
\begin{aligned}
G_{i,p} = \frac{m_{i,p} - rand}{m_{i=0,p} - rand}
\end{aligned}
\label{eq: norm}
\end{equation}
where $m_i,p$ is the probability of measuring the objective state, which is the solution to a factoring problem, and $rand$ is the probability of obtaining the correct solution from VQF by random selection. 

For example, if the number of variables to be solved is $4$, then the number of possible solutions is $2^4 = 16$. If there are $2$ correct solutions, then $rand$ will be $2/16$. The numerator term $m_{i,p} - rand$ can be thought of as the performance gain by applying VQF. In absence of noise (i.e. $i=0$), VQF can achieve full performance (i.e. $G_{i,p} = 100\%$). When the noise is large enough and $m_{i,p} \approx rand$, then $G_{i,p} \approx 0\%$ and applying VQF is equivalent to randomly guessing the answer. The motivation of integrating $rand$ into the calculation is to incorporate the sizes of instances into consideration, because the magnitude of $m_{i,p}$ can be problem instance dependent. For example, a $m_{i,p}$ with $15\%$ can be a very small performance gain for a small problem instance, where $rand$ is $12.5\%$, but can also be a decent solution for a large problem instance, where $rand$ is only $0.5\%$ in comparison. Therefore, by integrating $rand$ into the calculation, the evaluation will be more objective across different problem instances.

\begin{figure} [] 
\vspace{-1em}
 \begin{center}
    \includegraphics[width=0.5\textwidth]{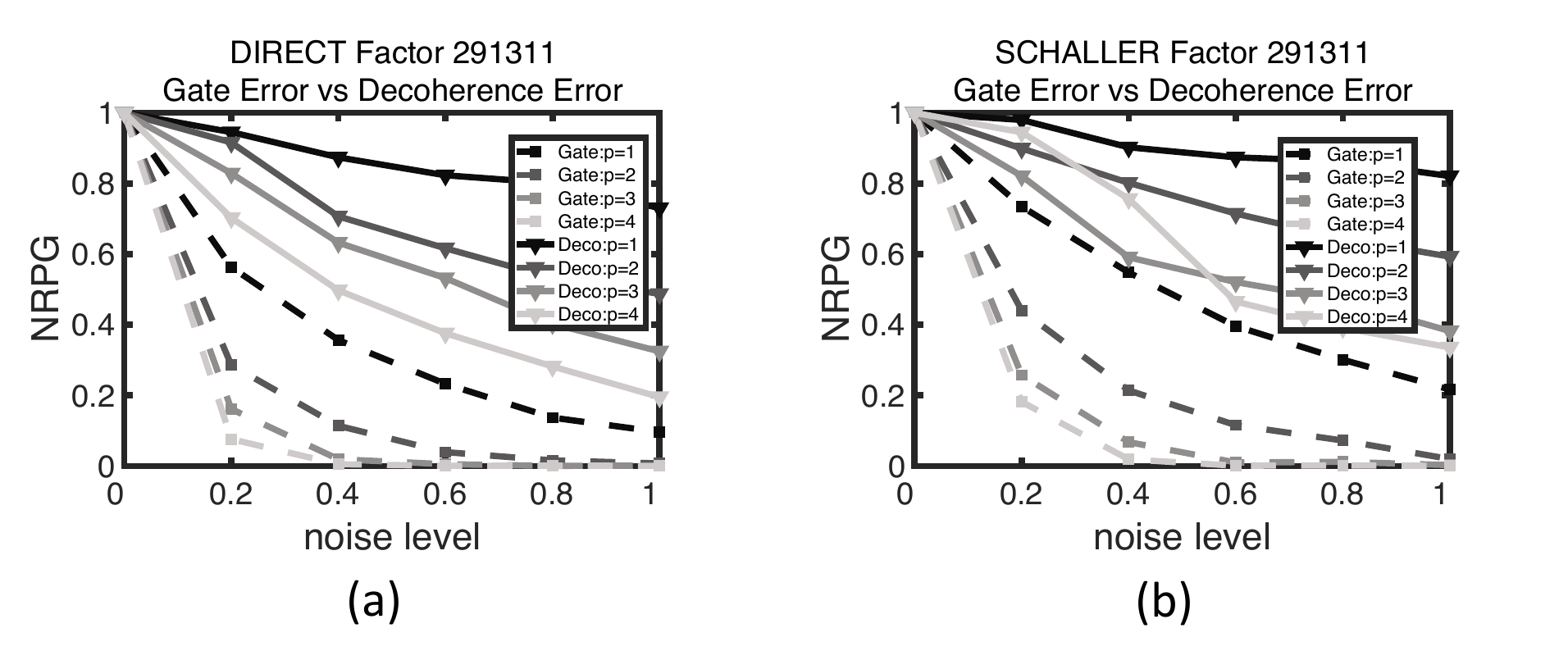}
 \end{center}
 \vspace{-1em}
 \caption{Comparison between gate noise (dashed lines) and decoherence noise (solid lines) for factoring 291311 using, (a) DIRECT and (b) SCHALLER transformation.} \label{fig:gvd}
 \vspace{-1mm}
\end{figure}


\begin{figure*} [t] 
\vspace{-1em}
 \begin{center}
    \includegraphics[width=0.98\textwidth]{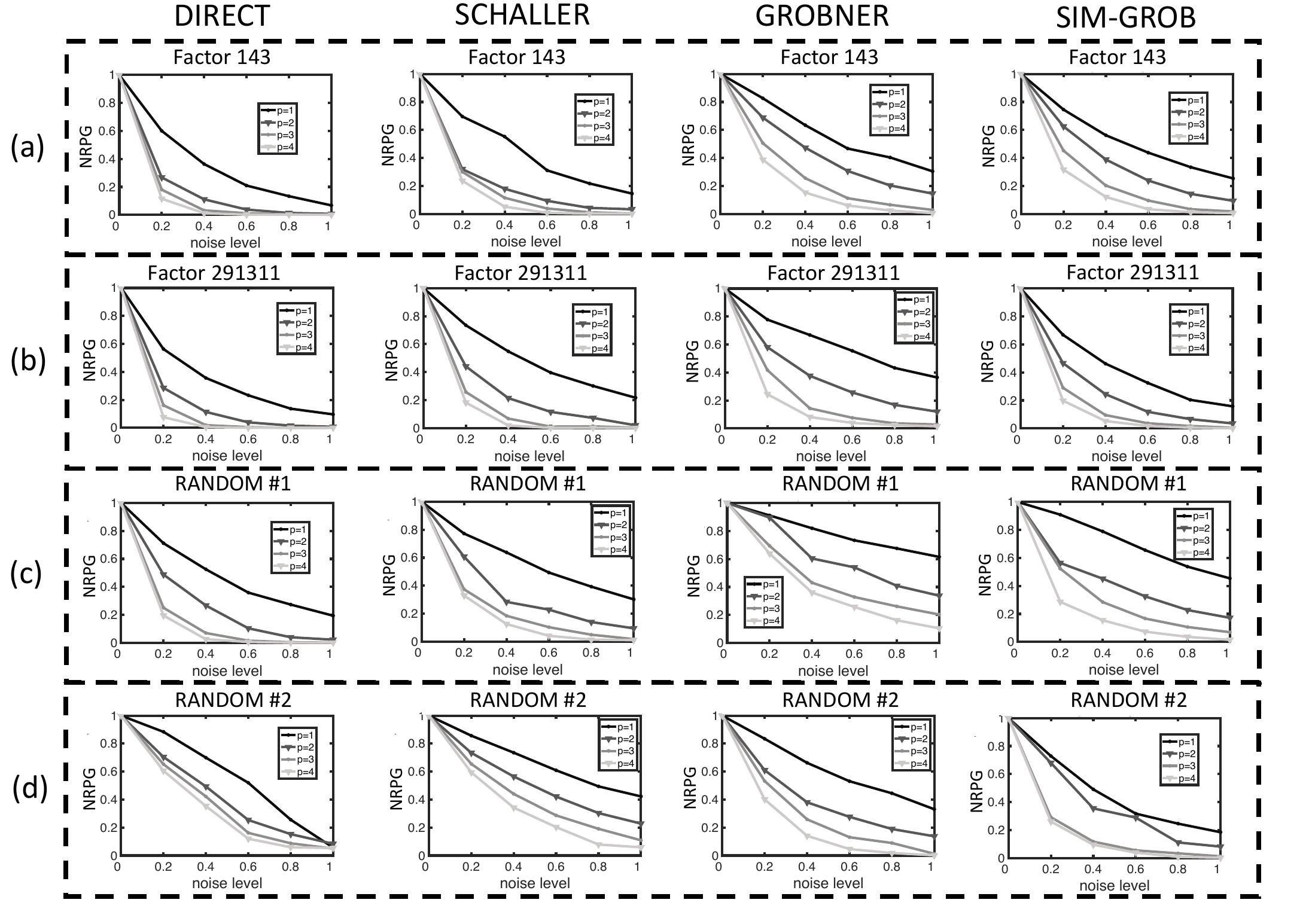}
 \end{center}
 \vspace{-1em}
 \caption{Comparison of transformation schemes under the impact of various noise levels for factoring, (a) 143; (b) 291311; (c) random Hamiltonian \#1; (d) random Hamiltonian \#2. } \label{fig:comparison}
 \vspace{-1mm}
\end{figure*}

\subsection{Impact of Gate Noise} 
\subsubsection{Gate Noise vs Decoherence} We first compare the impact of gate noise and decoherence noise on the performance of VQF (Fig. \ref{fig:gvd}). To solely examine the impact of gate noise, we mask decoherence noise and vice versa. For both factorization, the performance loss of VQF due to gate noise is much larger than that of decoherence noise. Therefore, the selection of quantum circuits for VQF should prioritize mitigating the impact of gate noise.

\subsubsection{Empirical Observation} In Fig. \ref{fig:comparison}, we demonstrate the error resiliency of various transformations for $4$ problem instances. Besides the analysis of solution space, we again note that the error resiliency of VQF is significantly impacted by the noise level for all the 4 transformations. The performance degradation increases with the noise level. Additionally, the QAOA level ($p$) can also impact the VQF performance. For a fixed noise level, increasing $p$ leads to more degradation due to noise. For example, for the GROBNER in Fig. 7(a), $G_{i=0.4,p}$ decreases from $63.37\%$ to $15.21\%$ when $p$ increases from $1$ to $4$. 

Transformations can also be a factor to determine the error resiliency of VQF under noise. For the first three numbers (i.e. $143$, $291311$ and Hamiltonian \#1), GROBNER provides relatively the best error resiliency; SIM-GROBNER is slightly better than the SCHALLER in some instances; DIRECT is the worst among all the schemes. We take factorization of $143$ as an example: when $i$ = 0.4 and $p=2$, the normalized performance gain for GROBNER, SIM-GROBNER, SCHALLER and DIRECT is: $47.26\%$, $38.79\%$, $17.91\%$ and $11.13\%$. However, for the last problem instance (Hamiltonian \#2), SIM-GROBNER has the lowest $G_{i,j}$ among all the transformation schemes, especially for $p=3$ and $4$. The reason can be inferred by comparing the circuit architectures in Fig. \ref{fig:arch}(d) i.e., SIM-GROBNER has significantly a larger number of CNOT gates and circuit depth than other schemes. 

\subsection{Selection of Resilient Quantum Circuit}
We can draw two conclusions from the previous observation: (a) the number of CNOT gates can greatly impact the noise resiliency of VQF. As indicated in Section III.D, DIRECT has more number of CNOT gates than GROBNER, SIM-GROBNER and SCHALLER for the first three problem instances. Thus, it has the worst error resiliency. For the last problem instance, SIM-GROBNER possess the largest number of CNOT gates. Correspondingly, it has the worst error resiliency among the transformation schemes; (b) the number of CNOT gate per qubit also impacts the noise resiliency of VQF. In Fig. \ref{fig:arch}(a)-(c), it can be noted that GROBNER, SCHALLER and SIM-GROBNER have a similar number of CNOT gates. However, since GROBNER and SIM-GROBNER have more qubits, their number of CNOT gates per qubit is lower than that of the SCHALLER. Therefore, GROBNER and SIM-GROBNER outperform SCHALLER in terms of the noise resiliency. This conclusion is further validated in the last problem instance, where GROBNER has more number of CNOT than DIRECT and SCHALLER, but their noise resiliency for this instance are about the same. This is due to higher number of qubits that can offset the more number of CNOT gates in GROBNER. However, one noteworthy tradeoff is that GROBNER and SIM-GROBER increase the number of qubits and may not be suitable for quantum computers with small number of qubits. 

Based on the above conclusions, we note that it is of great significance to choose an appropriate quantum circuit to improve the performance of VQF and that the selection can be done classically with very little timing overhead. The integration of the circuit selection stage into the workflow of VQF can be seen in Fig. 1.

\section{Discussions and Limitations}
\subsection{Future Transformation Techniques}
We have shown that gate noise is the dominant source that can impact the resiliency of VQF. We have identified two factors to alleviate this issue namely, usage of less number of CNOT gates or CNOT gates per qubit. Future work targeting new transformations guided by these two criteria can enhance the VQF resilience further.

\subsection{Considerations to Coupling Constraints}
We did not consider the coupling constraints of the quantum computers for the simplicity of analysis. Coupling constraints mandate two-qubit gate interaction between qubits that are connected physically. If not, one (or both) of the qubit(s) should be swapped to another qubit pair using the SWAP gates, with each consisting of $3$ CNOT gates. In other words, deploying a quantum circuit on a real quantum computer often will increase the number of CNOT gates and circuit depth due to the coupling constraints. 

The coupling constraints can be easily included in our proposed VQF flow by introducing the necessary SWAP gates. However, the developed selection criteria will remain valid for these quantum circuits.

\subsection{Validation with quantum hardware}
We considered the qubit quality metrics (e.g., gate error) for each qubit the same whereas they differ in reality. This assumption may have an effect in practice where one transformation can be benefited by better qubit allocation than others. However, our noise values are calibrated with real quantum hardware making the conclusions realistic. In the future work, we will implement the proposed VQF flow on real quantum computers and integrate more hardware-specific considerations to improve the accuracy of the analysis.

\section{Conclusions}
We analyzed the impact of noise on VQF for factoring both realistic and synthetic integers. We also explored 4 transformation techniques and their noise sensitivities (specifically to gate noise). We found that the quantum circuit structure resulted from transformation techniques can impact the performance of VQF i.e., the number of CNOT gates and the number of CNOT gates per qubit. Lastly, we improved the noise resiliency of VQF by integrating our findings into its workflow.

\bibliographystyle{IEEEtran}
\bibliography{IEEEabrv,biblio}

\end{document}